\documentclass[conference]{IEEEtran}
\IEEEoverridecommandlockouts
\usepackage{cite}
\usepackage{amsmath,amssymb,amsfonts}
\usepackage{algorithmic}
\usepackage{bbm}
\usepackage{graphicx}
\usepackage{diagbox}
\usepackage{textcomp}
\usepackage[font=small,labelfont=bf]{caption}
\usepackage{xcolor}
\usepackage{scalerel}
\usepackage{lettrine}
\usepackage{dsfont}
\usepackage{subcaption}
\usepackage{float}
\usepackage{amsmath,amsthm}
\usepackage[]{graphicx}
\usepackage{lipsum}
\def\BibTeX{{\rm B\kern-.05em{\sc i\kern-.025em b}\kern-.08em
    T\kern-.1667em\lower.7ex\hbox{E}\kern-.125emX}}
\begin{document}

\title{Space Diversity-Based Grant-Free Random Access for Critical and Non-Critical IoT Services\\
\thanks{The work of Rahif Kassab and Osvaldo Simeone has received funding from the European  Research  Council (ERC) under the European Union Horizon 2020 research and innovation program (grant agreement 725731).}
}

\author{\IEEEauthorblockN{Rahif Kassab\IEEEauthorrefmark{1}, Osvaldo Simeone\IEEEauthorrefmark{1}, Andrea Munari\IEEEauthorrefmark{2} and Federico Clazzer\IEEEauthorrefmark{2}  }\\
\IEEEauthorblockA{\small\IEEEauthorrefmark{1}King's Centre for Learning and Information Processing (kclip), King's College London, London, UK\\
\small\IEEEauthorrefmark{2} Institute of Communications and Navigation, German Aerospace Center (DLR),  82234 Wessling, Germany\\
Emails: \IEEEauthorrefmark{1}\{rahif.kassab,osvaldo.simeone\}@kcl.ac.uk , \IEEEauthorrefmark{2}\{Andrea.Munari,Federico.Clazzer\}@dlr.de}}

\maketitle

\begin{abstract}
In this paper, we study the coexistence of critical and non-critical Internet of Things (IoT) services on a grant-free channel consisting of radio access and backhaul segments. On the radio access segment, IoT devices send packets to access points (APs) over an erasure collision channel using the slotted ALOHA protocol. Then, the APs forward correctly received messages to a base station (BS) over a shared wireless backhaul segment, modeled as an erasure collision channel. The APs hence play the role of uncoordinated relays that provide space diversity and may reduce performance losses caused by collisions. Both non-orthogonal and inter-service orthogonal resource sharing are considered and compared. Throughput and reliability metrics are analyzed, and numerical results are provided to assess the performance trade-offs between critical and non-critical IoT services.
\end{abstract}
\begin{IEEEkeywords}
Beyond 5G, IoT, Grant-Free, Radio Access
\end{IEEEkeywords}
\section{Introduction}
Future generations of cellular and satellite networks, starting with 5G, will cater to heterogeneous services with vastly different performance requirements \cite{5Goverview}\cite{opportunistic_coexistence}. Among these services are Internet of Things (IoT) networks characterized by short and sporadic packet transmissions, which will support applications with critical or non-critical requirements in terms of reliability. \par
\begin{figure}[h]
	\centering
	\includegraphics[height= 7.5 cm, width= 8.5 cm]{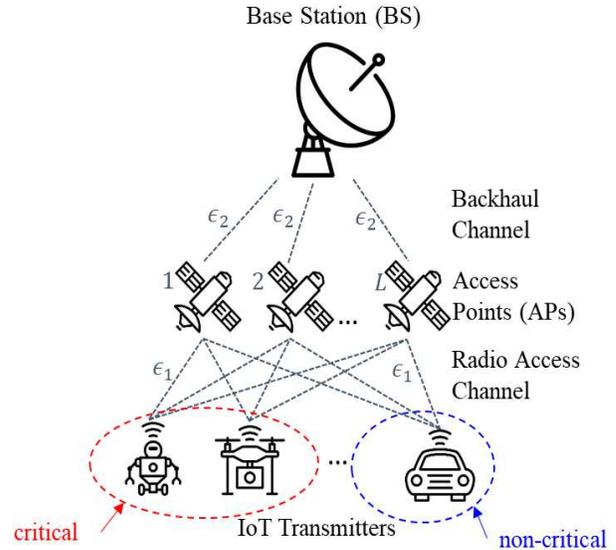}
	\caption{An IoT system with grant-free wireless radio access and shared backhaul with uncoordinated APs, in which IoT devices generate critical or non-critical messages. The set-up in the figure illustrates a special instance of the model with APs as LEO satellites and BS as ground station.}
	\label{fig:system_model}
\end{figure}
In the presence of a large number of IoT devices such as in massive Machine Type Communications (mMTC) scenarios \cite{mmtcsaad}, conventional grant-based radio access protocols can cause a significant overhead on the access network due to the large number of handshakes to be established. A potentially more efficient solution is given by grant-free radio access protocols, which are used by many commercial solutions both in the terrestrial domain, e.g. Sigfox \cite{sigfox} and LoRa \cite{lora} and in the satellite domain, using constellations of Low-Earth Orbit (LEO) satellites to collect information, e.g., Orbcomm \cite{orbcomm} and Myriota \cite{myriota}. Under grant-free access, devices transmit whenever they have a packet to deliver without any prior handshake \cite{grant_free_popovski,rahif_grant_free,grant_free_cavdar}. This is typically done via some variants of the classical ALOHA random access scheme \cite{abramson1970aloha}. \par 
In the presence of different IoT services and devices, orthogonal resource allocation schemes such as inter-service Time Division Multiple Access (TDMA) are used \cite{3gpp_nbiot}. Orthogonal schemes may cause an inefficient use of resources in future IoT scenarios due to limited spectral resources and inherent inefficiency when traffic patterns become unpredictable. Recent work has hence proposed to apply non-orthogonal resource allocation to heterogeneous services \cite{rahif_access_2018}\cite{popovski2018slicing}. In order to mitigate interference in non-orthogonal schemes, one can leverage successive interference cancellation (SIC) \cite{aloha_noma}, time diversity \cite{coded_slotted_aloha}, and/or space diversity \cite{munari_multiple_aloha}\cite{vladimir_cooperative_ALOHA}. The latter is provided by multiple Access Points (APs) that play the role of relays between the devices and the Base Station (BS), as illustrated in Fig.~\ref{fig:system_model}.\par 
In this work, we study the space diversity-based model introduced in Fig.~\ref{fig:system_model} that provides grant-free access to both critical and non-critical services. We assume uncoordinated APs, so that both radio access and backhaul channels are operating using ALOHA. The lack of coordination among APs can be considered as a worst-case analysis for dense low-cost cellular deployments \cite{het_networks_no_coordination} \cite{vladimir_cooperative_ALOHA}. It also may account for the scenario in Fig.~\ref{fig:system_model}, where a constellation of LEO satellites act as relays between ground terminals and a central ground station, since the presence of inter-satellite links is too costly to be deployed. For the system in Fig.~\ref{fig:system_model}, we derive throughput and reliability measures for critical and non-critical services as a function of key parameters such as the number of APs and traffic loads. The analysis accounts for orthogonal and non-orthogonal inter-service protocols and considers two receiver models, namely, \textit{superposition} and \textit{collision models} as detailed in the next section. The most related prior work is \cite{frederico2019modern}, in which a simplified collision model with only a single service was considered for the same space-diversity model.\par
The rest of the paper is organized as follows. In Sec.~\ref{sec:system_model_performance_metrics} we describe the system model used and the performance metrics. In Sec.~\ref{sec:throughput_reliability}, we derive throughput and reliability under the erasure channels model. Numerical results are provided in Sec.~\ref{sec:numerical_results}, and conclusions are drawn in Sec.~\ref{sec:conclusions}. 
\section{System model and Performance metrics}
\label{sec:system_model_performance_metrics}
\subsection{System Model}
We consider the system illustrated in Fig.~\ref{fig:system_model}, in which $L$ APs, e.g., LEO satellites, provide connectivity to IoT devices. The APs are in turn connected to a BS, e.g., a ground station, through a shared wireless backhaul channel. We assume that time over both access and backhaul channels is divided into frames and each frame contains $T$ time-slots. At the beginning of each frame, a random number of IoT devices are active. The number of active IoT devices that generate critical and non-critical messages at the begining of the frame follow independent Poisson distributions with average loads $\gamma_c G$ and $(1-\gamma_c)G\ \mathrm{[packet/slot]}$, respectively, for some parameter $\gamma_c \in [0,1]$ and total system load $G$. Users select a time-slot $t$ uniformly at random among the $T$ time-slots in the frame and independently from each other. By the Poisson thinning property \cite{billingsley2008probability}, the random number $N_c(t)$ of critical messages transmitted in a time-slot $t$ follows a Poisson distribution with average $G_c = \gamma_c G /T$, while the random number $N_{\bar{c}}(t)$ of non-critical messages transmitted in slot $t$ follows a Poisson distribution with average $G_{\Bar{c}}=(1-\gamma_c)G/T$.\par
\textit{Radio Access Model:} As in, e.g., \cite{frederico2019modern,azimi2017content,calderbank_erasure}, we model the access links between any device and an AP as an independent interfering erasure channel with erasure probability $\epsilon_1$. Specifically, a packet sent by a user is independently erased at each receiver with probability $\epsilon_1$, causing no interference, or is received with full power with probability $1-\epsilon_1$. The erasure channels are independent and identically distributed (i.i.d.) across all slots and frames. Interference from messages of the same type received at an AP is assumed to cause a destructive collision. Furthermore, critical messages are assumed transmitted with a higher power than non-critical messages so as to improve their reliability, hence creating significant interference on non-critical messages. As a result, in each time-slot, an AP can be in three possible states:\\
$\bullet$ a critical message is retrieved successfully if the AP receives only one critical message. Critical messages are assumed to be immune to non-critical transmissions due to their large transmission power;\\
$\bullet$ a non-critical message is retrieved if the AP receives only one non-critical message and zero critical messages;\\
$\bullet$ no message is retrieved if multiple critical messages and/or non-critical messages are received at the AP, or if no messages are received due to channel erasures, or also if no messages were transmitted (i.e., none of the devices is active).\par
\textit{Backhaul model:} The APs share a wireless out-of-band backhaul that operates in a full-duplex mode and in an uncoordinated fashion as in \cite{frederico2019modern}.
In each time-slot $t+1$, an AP sends a message retrieved on the radio access channel in the corresponding time-slot $t$ to the BS over the backhaul channel. APs with no message retrieved in slot $t$ remain silent in the corresponding backhaul time-slot $t+1$. The link between each AP and the BS is modeled as an erasure channel with erasure probability $\epsilon_2$, and destructive collisions occur at the BS if two or more messages of the same type are received. As for the radio access case, erasure channels are i.i.d. across APs, slots and frames.\par
In order to model interference between APs, we consider two scenarios. The first, referred to as \textit{collision model}, assumes that multiple messages from the same device cause destructive collision. Under this model, in each time-slot, the BS's receiver can be in three possible states:\\
$\bullet$ a critical message is retrieved successfully at the BS is only one critical message is received. As in the radio access scenario, critical messages are not affected by non-critical messages due to their larger transmission power;\\
$\bullet$ a non-critical message is retrieved successfully if no other critical or non-critical message is received;\\
$\bullet$ no message is retrieved at the BS if multiple critical or non-critical messages are received at the BS or no messages are received due to channel erasures or also no messages were transmitted.\par
In the second model, referred to as \textit{superposition model}, the BS is able to decode from the superposition of multiple instances of the same packet that are relayed by different APs on the same backhaul slot, assuming no other transmission occured on it. In practice, this can be accomplished by ensuring that the time asynchronism between APs is no larger that the cyclic prefix in a multicarrier modulation implementation. This can be done, for example, by having a central master clock at the BS against which the local time bases of APs are synchronized \cite{timesynchro_patent_AP}. Note that this model is valid for uncoordinated APs. Hence, the BS's receiver can be in three possible states:\\
$\bullet$ a critical message is retrieved successfully at the BS in a given time-slot if no \textit{different} critical message is received by the BS;\\
$\bullet$ a non-critical message is retrieved successfully if no critical messages and no \textit{different} non-critical messages are received in the same slot;\\
$\bullet$ no message is retrieved at the BS if multiple different critical or non-critical messages are received at the BS or no message is received due to channel erasures, or also if no messages were transmitted. 
\par In addition to non-orthogonal resource allocation whereby devices from both services share the entire frame of $T$ time-slots, we also consider orthogonal resource allocation, namely \textit{inter-service time division multiple access} (TDMA) where a fraction $\alpha T$ of the frame's time-slots are reserved to critical devices and the remaining $(1-\alpha)T$ for non-critical devices. Inter-service contention in each allocated fraction follows a slotted ALOHA protocol as discussed above. In the following, we derive the performance metrics under the more general non-orthogonal scheme described above. The performance metrics under TDMA for each service can be obtained by replacing $T$ with the corresponding fraction of resources in the performance metrics equations and taking the interference from the other service to zero.
\subsection{Performance Metrics}
\label{sec:performance_metrics}
We are interested in computing the throughputs $R_c$ and $R_{\bar{c}}\ [\mathrm{packet/slot}]$ and the reliability levels $\Gamma_c$ and $\Gamma_{\bar{c}}\ [\mathrm{packet/frame}]$ for critical and non-critical messages respectively. The throughputs are defined as the average number of packets decoded correctly in any given time-slot at the BS for each type of service. The reliability levels are defined by the average fraction of critical and non-critical packets generated in a frame that are retrieved by the BS by the end of the frame.
\section{Throughput and Reliability Analysis}
\label{sec:throughput_reliability}
In this section, we derive the throughputs and reliability levels for both types of messages under the collision and superposition models described above. Throughout the discussion, we denote as $X\sim \operatorname{Bin}(n,p)$ a Binomial random variable (RV) with $n$ trials and probability of success $p$; as $X ~\sim \operatorname{Poiss}(\lambda)$ a Poisson RV with parameter $\lambda$. We also write $(X,Y)\sim f \cdot g$ for two independent RVs $X$ and $Y$ with respective probability density functions $f$ and $g$.
\subsection{Collision Model}
Under the collision model, two messages received at the BS in the same time-slot and generated from the same devices undergo a destructive collision. We start by introducing RVs $B_i(t)$ for the state of the $i$-th AP, with $i=1,\ldots,L$ and RV $B(t)$ for the state of the BS in any time-slot $t=1,\ldots,T$. Since all RVs are i.i.d. across time-slots, the index $t$ is dropped for simplicity of notation whenever no confusion may arise. These RVs take values as
\begin{equation}
    B_i \sim B = \begin{cases}
    c & \text{if a critical message is retrieved}  \\
    \bar{c} &  \text{if a non-critical message is retrieved} \\
    0 & \text{if no message is retrieved due to erasures} \\ & \text{or collisions or no transmitted messages}
    \end{cases} \label{eq:B_collision}
\end{equation}
in the given time-slot and for $i=1,\ldots,L$. Furthermore, we denote by $M_c$ and $M_{\Bar{c}}$ the RVs representing the overall number of received critical and non-critical messages, respectively, at all the APs in a given time-slot. Accordingly, RVs $M_c$ and $M_{\bar{c}}$ can be written as
\begin{equation}
    M_c = \sum_{i=1}^{L} \mathbbm{1}_{ \{ B_i = c \}}
\ \ \text{and}\ \ 
     M_{\bar{c}} = \sum_{i=1}^{L} \mathbbm{1}_{ \{ B_i = \bar{c} \}}.
\end{equation}
where $\mathbbm{1}_{\{ a\}}$ is the indicator function of an event $a$.
Conditioned on the number of transmitted messages $N_c$ and $N_{\Bar{c}}$, RVs $M_c$ and $M_{\bar{c}}$ are distributed as
\begin{equation}
    M_{\bar{c}}|N_c,N_{\bar{c}} \sim \operatorname{Bin}(L,p_{\bar{c}}) \label{eq:dis_Mbarc_collision}
\end{equation}
and 
\begin{equation}
    M_c|M_{\bar{c}},N_c,N_{\bar{c}} \sim \operatorname{Bin}(L-M_{\bar{c}},p_c), \label{eq:dis_Mc_collision}
\end{equation}
with the corresponding parameters given as
\begin{subequations}
\begin{alignat}{1}
& p_{c}\! =\! \mathrm{Pr}[B_i = c|N_c\! = \!n_c, N_{\Bar{c}}\!=\!n_{\bar{c}} ]=  n_c (1\!-\!\epsilon_1)\epsilon_1^{n_c -1} \label{eq:pc}\\
    \mathrm{and}\ &  p_{\Bar{c}}\! =\!\mathrm{Pr}[B_i \!=\!\Bar{c}| N_c=n_c,N_{\Bar{c}}=n_{\bar{c}}] =  n_{\Bar{c}} (1 \! - \! \epsilon_1) \epsilon_1^{n_{\Bar{c}}-1} \epsilon_1^{n_c}. \label{eq:pbarc}
\end{alignat} \label{eq:pc_and_pbarc}
\end{subequations}
The expression \eqref{eq:pc} is the probability of an AP receiving a critical message from any of the $N_c=n_c$ active critical devices in the slot. The expression \eqref{eq:pbarc} is the probability of an AP receiving a non-critical message from any of the $N_{\bar{c}}=n_{\bar{c}}$ active non-critical devices. Note that the latter requires all critical messages to be erased which is represented by the probability term $\epsilon_{1}^{n_c}$.
\par
Following a similar reasoning, given $M_c$, $M_{\bar{c}}$, $N_c$ and $N_{\bar{c}}$, the probability of retrieving successfully a critical message at the BS in a given time-slot can be written as
\begin{equation}
\begin{aligned}
    q_c& =\mathrm{Pr}[B=c|N_c = n_c , N_{\bar{c}}=n_{\bar{c}} , M_c = m_c, M_{\bar{c}}=m_{\bar{c}}] \\ & = m_c (1 - \epsilon_2)\epsilon_2^{m_c - 1}. \label{eq:m_c_Bin}
    \end{aligned}
\end{equation}
The probability of retrieving a non-critical message at the BS is given as 
\begin{equation}
\begin{aligned}
    q_{{\bar{c}}} & = \mathrm{Pr}[B=\bar{c} | N_c = n_c , N_{\bar{c}}=n_{\bar{c}} , M_c = m_c, M_{\bar{c}} = m_{\bar{c}}] \\ & =m_{\bar{c}} (1 - \epsilon_2) \epsilon_2^{m_c} \epsilon_2^{m_{\Bar{c}}- 1} .
    \end{aligned}
\end{equation}
Removing the conditioning on $M_c,M_{\Bar{c}},N_c$ and $N_{\Bar{c}}$ and using the  distributions \eqref{eq:dis_Mbarc_collision} and \eqref{eq:dis_Mc_collision}, the throughputs can be directly computed as the expectations
\begin{equation}
    R_c=\mathbb{E}[q_c]\ \ \text{and}\ \  
    R_{\bar{c}}=\mathbb{E}[ q_{\Bar{c}}] \label{eq:expectation_rate}
\end{equation}
where averages are taken over RVs $N_c, N_{\bar{c}}, M_c$ and $M_{\bar{c}}$. These expectations can be derived in closed form as detailed in \cite{rahif_uncoordinated}.
\par Given the above definitions, the reliability levels of critical and non-critical messages can be written respectively as
\begin{subequations}
\begin{alignat}{1}
    & \Gamma_c = \mathbb{E}\Bigg[ \frac{\sum_{t=1}^{T} \mathbbm{1}_{\{B(t)=c \}}}{\sum_{t=1}^{T} N_c (t)} \bigg| \sum_{t=1}^{T} N_c(t) \geq 1 \Bigg] \label{eq:reliability_collision_a} \\ &\text{and}\ \   \Gamma_{\bar{c}}= \mathbb{E}\Bigg[ \frac{\sum_{t=1}^{T} \mathbbm{1}_{\{B(t)=\bar{c} \}}}{\sum_{t=1}^{T} N_{\bar{c}}(t)} \bigg| \sum_{t=1}^{T} N_{\bar{c}}(t) \geq 1 \Bigg],
    \label{eq:reliability_collision_b}\end{alignat} \label{eq:reliability_collision}
\end{subequations}
\! \! \! \!\!\!\!\!with expectations taken over RVs $N_c(t), N_{\bar{c}}(t), M_c(t)$, $M_{\bar{c}}(t)$, and $B(t)$ across all slots $t=1, \ldots, T$. The conditioning in \eqref{eq:reliability_collision} ensures that at least one packet of the given type is transmitted in the given frame. The conditional joint distributions needed to compute \eqref{eq:reliability_collision_b} are defined through the chain rule by the distributions
\begin{subequations}
\begin{alignat}{1}
   & \{N_c(t),N_{\bar{c}}(t) \}_{t=1}^{T} \bigg| \sum_{t=1}^{T} N_c(t) \geq 1 \sim \nonumber \\ & \ \ \ \Big( \prod_{t=1}^{T}  \text{Poiss}(n_c | g_c) \Big) \Big( \frac{1}{Z} \prod_{t=1}^{T}\text{Poiss}(n_{\bar{c}} | g_{\bar{c}}) \mathbbm{1}_{\{\sum_{t=1}^{T} N_{\bar{c}}(t) \geq 1 \}} \Big) \\
     &\ \mathrm{and}\ \  \{M_c(t), M_{\bar{c}}(t)  \}_{t=1}^{T} \bigg| \{N_c(t),N_{\bar{c}}(t) \}_{t=1}^{T} \sim \nonumber \\ & \ \ \ \ \ \ \ \ \ \ \ \ \ \prod_{t=1}^{T} \operatorname{Bin}(L, p_{\bar{c}}(t)) \operatorname{Bin}(L-M_{\bar{c}} (t), p_{{c}}(t)),
\end{alignat} 
\end{subequations}
where $Z = 1 - \mathrm{Pr}(\sum_{t=1}^{T} N_{\bar{c}} (t)= 0)$ is a normalizing factor; and $p_c(t)$ and $p_{\bar{c}}(t)$ are defined as in \eqref{eq:pc_and_pbarc} with $N_c(t)$ and $N_{\bar{c}}(t)$ in lieu of $n_c$ and $n_{\bar{c}}$, respectively.
Similar expressions apply for \eqref{eq:reliability_collision_a}. Note that, conditioned on there being at least one non-critical message transmitted in the frame, the RVs $\{N_{\bar{c}}(t) \}_{t=1}^{T}$ are not i.i.d.
\subsection{Superposition Model}
In this subsection, we derive the throughput and reliability measures of critical and non-critical messages under the superposition model. To this end, unlike for the collision model, one needs to keep track of the index of the messages decoded by the APs in order to be able to detect when multiple instances of the same message (i.e., sent by the same device) are received at the BS. We start by defining the RVs $B_i$ to denote the index of the message received at AP $i$ and RV $B$ for the BS at any time-slot. Accordingly, for given values $N_c=n_c$ and $N_{\bar{c}}=n_{\bar{c}}$ of transmitted messages, RVs $\{B_i\}$ can take values
\begin{equation}
    \begin{cases}
    0 & \text{if no message is retrieved due to}\\ & \text{erasures or collisions }\\
    1 \leq m \leq n_c & \text{if the $m$-th critical message is }\\ & \text{retrieved}  \\
    n_c + 1 \leq m \leq n_c + n_{\bar{c}} &  \text{if the $(m-n_c)$-th non-critical }\\ &\text{message is retrieved.}
 \label{eq:Bi_superposition}
    \end{cases}
\end{equation}
Note that we have indexed critical messages from $1$ to $n_c$ and non-critical messages from $n_c + 1$ to $n_c + n_{\bar{c}}$. $B$ is defined as in \eqref{eq:B_collision}. Furthermore, we define as $M_m = \sum_{i=1}^{L} \mathbbm{1}_{\{B_i=m\}}$ the RVs denoting the number of APs that have message of index $m \in \{0,1, \ldots, n_c,n_c+1, \ldots , n_c + n_{\bar{c}} \}$. The joint distribution of RVs $\{M_m \}_{m=0}^{n_c+n_{\bar{c}}}$ given $N_c$ and $N_{\bar{c}}$ is multinomial and can be written as follows
\begin{equation}
\begin{aligned}
    & \{M_m \}_{m=0}^{n_c+n_{\bar{c}}}|N_c,N_{\bar{c}} \sim \\&\operatorname{Multinomial}\Big(L ,\overbrace{1-p_c-p_{\bar{c}}}^{0},\overbrace{\frac{p_c}{n_c}, \ldots,\frac{p_c}{n_c}}^{n_c}, \overbrace{\frac{p_{\bar{c}}}{n_{\bar{c}}},\ldots, \frac{p_{\bar{c}}}{n_{\bar{c}}}}^{n_{\bar{c}}} \Big), \label{eq:multinomial}
    \end{aligned}
\end{equation}
where we used the the probabilities in \eqref{eq:pc_and_pbarc} that one of the critical or non-critical message is received at an AP respectively in a given time-slot. 
The probability of retrieving a critical message in a given time-slot at the BS conditioned on $N_c$, $N_{\bar{c}}$ and $\{M_{m\prime} \}_{m\prime=0}^{n_c + n_{\bar{c}}}$ can be then written as 
\begin{equation}
\begin{aligned}
    q_c& =\mathrm{Pr}[B=c | N_c=n_c , N_{\bar{c}}=n_{\bar{c}}, \{M_{m^\prime} \}_{m^\prime=0}^{n_c + n_{\bar{c}}}]   \\ & = \sum_{m=1}^{n_c} \sum_{j=1}^{M_m} {M_m \choose j} (1-\epsilon_2)^j \epsilon_2^{\delta_1}, \label{eq:proba_c_superposition} 
    \end{aligned}
\end{equation}
where $\delta_1$ is defined as follows
\begin{equation}
    \delta_1= \sum_{ \substack{m^\prime = 0 \\ m^\prime \neq m}}^{n_c} M_{m^\prime} + M_m - j. \label{eq:delta_1}
\end{equation}
The first sum in \eqref{eq:proba_c_superposition} is over all possible critical messages and the second sum is over all combinations of APs that have the critical message $m$. The sum in \eqref{eq:delta_1} is over all APs that have a critical message $m^{\prime} \neq m$. The throughput of critical messages can be computed by averaging \eqref{eq:proba_c_superposition} over all conditioning variables as
\begin{equation}
    R_c = \mathbb{E}_{N_c , N_{\bar{c}} , \{M_m\}_{m=0}^{N_c+N_{\bar{c}}}} [q_c].
\end{equation}
In a similar manner, the conditional probability of receiving a non-critical message at the BS can be written as
\begin{equation}
\begin{aligned}
    & q_{\bar{c}} =\mathrm{Pr}[B = \bar{c} |  N_c=n_c , N_{\bar{c}}=n_{\bar{c}} , \{M_{m^\prime} \}_{m^\prime = 0}^{n_c + n_{\bar{c}}} ] \\ &=  \sum_{m=n_c  +  1}^{n_c + n_{\bar{c}}}  \sum_{j=1}^{M_m} {M_m \choose j} (1-\epsilon_2)^j \epsilon_2^{\delta_2},
    \end{aligned} \label{eq:proba_cbar_one}
\end{equation}
where $\delta_2$ is written as
\begin{equation}
    \delta_2 = \sum_{ \substack{m^{\prime\prime}=n_c + 1 \\ m^{\prime\prime} \neq m } }^{n_c + n_{\bar{c}}} M_{m^{\prime \prime}} + M_m  - j +\sum_{m^\prime = 1}^{n_c} M_{m^\prime}. \label{eq:delta_2}
\end{equation}
The first sum in \eqref{eq:proba_cbar_one} is over all possible non-critical messages $m$ while the second sum is over all possible combinations of APs that have message $m$. The first and second sums in \eqref{eq:delta_2} are over all APs that have a different non-critical message and a critical message respectively. The throughput of non-critical messages can be then obtained by averaging over the conditioning RVs as
\begin{equation}
    R_{\bar{c}} = \mathbb{E}_{N_c, N_{\bar{c}}, \{ M_m \}_{m= 0}^{N_c + N_{\bar{c}}} } [q_{\bar{c}}].
\end{equation}
\par The reliability levels under the superposition model can be defined as in \eqref{eq:reliability_collision} with the caveat that one needs to average over the RVs $M_m(t)$, for $m\in \{0,1,\ldots , n_c + n_{\bar{c}} \}$ and $t =  1 , \ldots, T$ instead of $M_c (t)$, by using the distribution in \eqref{eq:multinomial}.
\section{numerical Results}
\label{sec:numerical_results}
\begin{figure}[h]
	\centering
	\includegraphics[height= 6.3 cm, width= 9.5 cm]{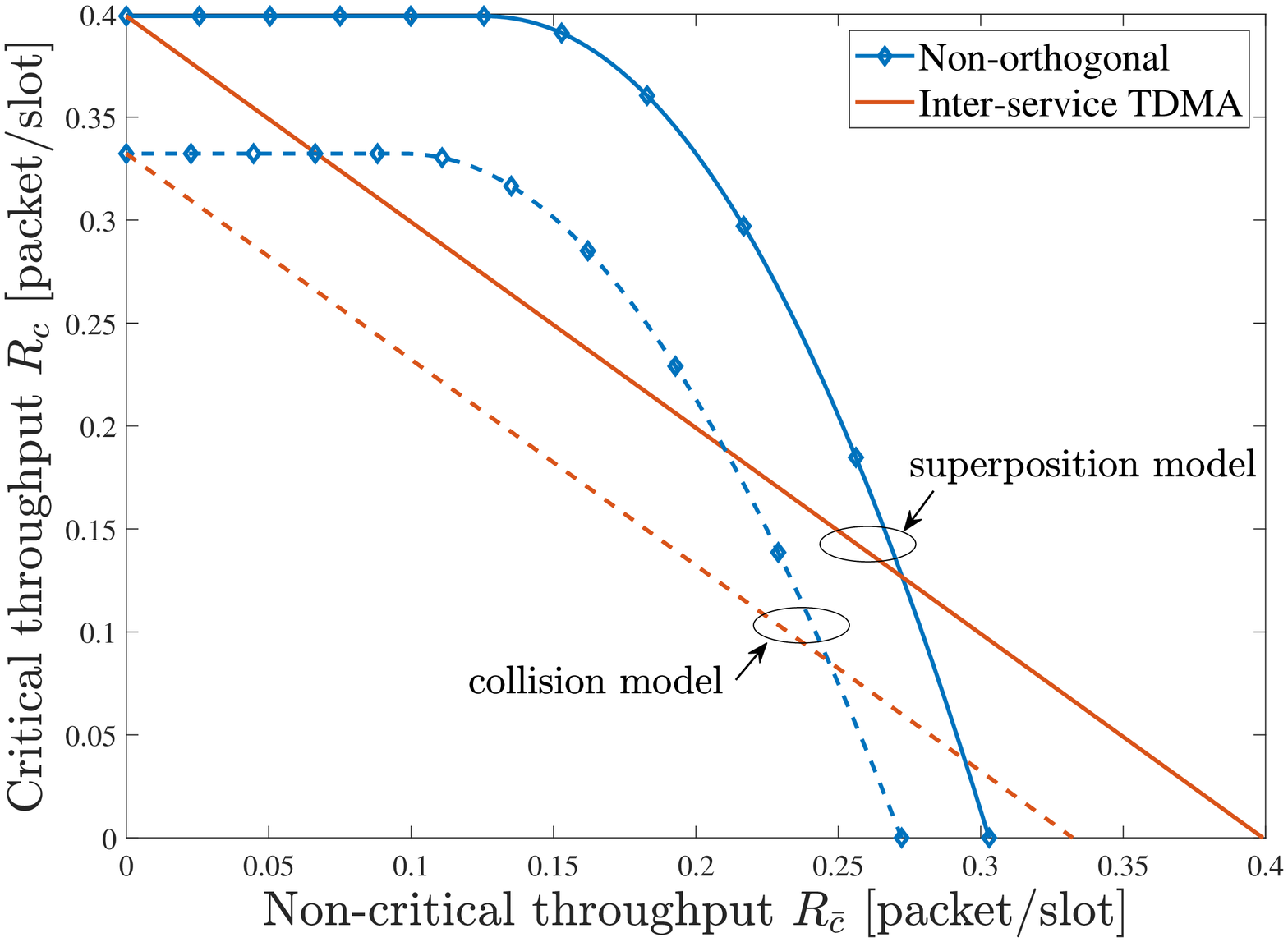}
	\caption{Achievable throughput region for critical and non-critical services for superposition and collision models ($\epsilon = 0.5$, $G=16\ [\mathrm{packet/frame}]$, $T=4\ \mathrm{[time\textrm{-}slot/frame]}$, and $L = 3$ APs).}
	\label{fig:rate_region}
	\vspace{-10pt}
\end{figure}
In this section, we numerically evaluate performance trade-offs in terms of throughput and reliability level for both services as function of key system parameters such as the channel erasure probabilities $\epsilon_1$ and $\epsilon_2$, number of APs $L$, and frame duration $T$. Unless specified otherwise, we assume throughout this section that we have $\epsilon_1=\epsilon_2=\epsilon$.\par
We start by plotting the region of achievable throughputs for critical and non-critical messages for both collision and superposition models in Fig.~\ref{fig:rate_region} for $\epsilon = 0.5$, total load $G=16\ [\mathrm{packet/frame}]$, $T=4\ \mathrm{[time\textrm{-}slot/frame]}$, and $L = 3$ APs. The region includes all throughput pairs that are achievable for some value of the fraction $\gamma_c$ of critical messages, as well as all throughput pairs that are dominated by an achievable throughput pair (i.e., for which both critical and non-critical throughputs are smaller than for an achievable pair). For reference, we also plot the throughput region for a conventional inter-service TDMA protocol, whereby a fraction $\alpha T$ for $\alpha \in [0,1]$ of the $T$ time-slots is allocated for critical messages and the remaining time-slots to non-critical messages. For TDMA, the throughput region includes all throughput pairs that are achievable for some value of $\alpha$, as well as of $\gamma_c$. A first observation from the figure is that non-orthogonal resource allocation can accommodate a significant non-critical throughput without affecting the critical throughput, while TDMA causes a reduction in the critical throughput for any increase in the non-critical throughput. This is due to the need in TDMA to allocate orthogonal time resources to non-critical messages in order to increase the corresponding throughput. However, with non-orthogonal resource allocation, the maximum non-critical throughput is generally penalized by the interference caused by the collisions from critical messages, while this is not the case for TDMA. In brief, TDMA is preferable when one wishes to guarantee a large non-critical throughput and the critical throughput requirements are loose; otherwise, non-orthogonal resource allocation outperforms TDMA in terms of throughput. Finally, we observe that significant gains can be obtained under the superposition model, leveraging as useful the superposition of multiple packets containing the same message.\par
\begin{figure}[h]
	\centering
	\includegraphics[height= 6.3 cm, width= 9.5 cm]{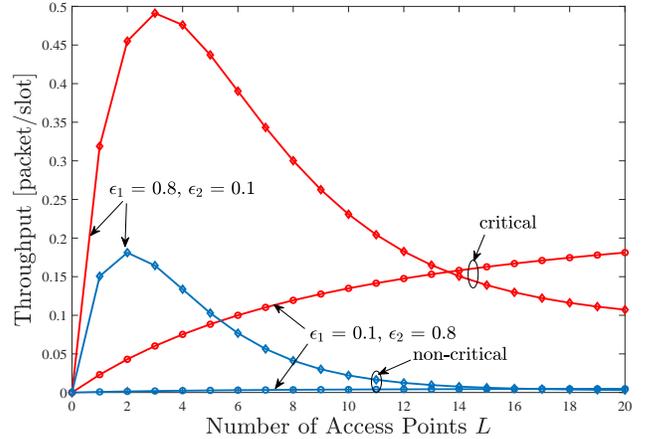}
	\caption{Critical and non-critical throughputs as function of the number of APs $L$ under the superposition model under non-orthogonal resource allocation ($G=30\ [\mathrm{packet/frame}]$, $T=4\ \mathrm{[time\textrm{-}slot/frame]}$, $\gamma_c = 0.5$ and for $\epsilon_1 \neq \epsilon_2$).}
	\vspace{-14pt}
	\label{fig:rate_L}
\end{figure}
In Fig.~\ref{fig:rate_L}, we explore the effect of the number of APs $L$ on the throughputs of both type of messages. To capture separately the effects of the radio access and the backhaul channel erasures, we consider here different values of the channel erasure probabilities $\epsilon_1$ and $\epsilon_2$. We highlight two different regimes: the first is when $\epsilon_1$ is large and $\epsilon_2$ is small, and hence larger erasures occur on the access channel; while the second covers the complementary case where $\epsilon_1$ is small and $\epsilon_2$ is large. In the first regime, increasing the number of APs is initially beneficial to both critical and non-critical messages in order to provide additional spatial diversity for the radio access given the large value of $\epsilon_1$; but larger values of $L$ eventually increase the probability of collisions at the BS on the backhaul due to the low value of $\epsilon_2$. In the second regime, when $\epsilon_1=0.1$ and $\epsilon_2=0.8$ much lower throughputs are obtained due to the significant losses on the backhaul channel. This can be mitigated by increasing the number of APs, which increases the probability of receiving a packet at the BS.
\begin{figure*}
\centering
\begin{subfigure}{.5\textwidth}
  \centering
  \includegraphics[height=5.5 cm, width= 7.5 cm]{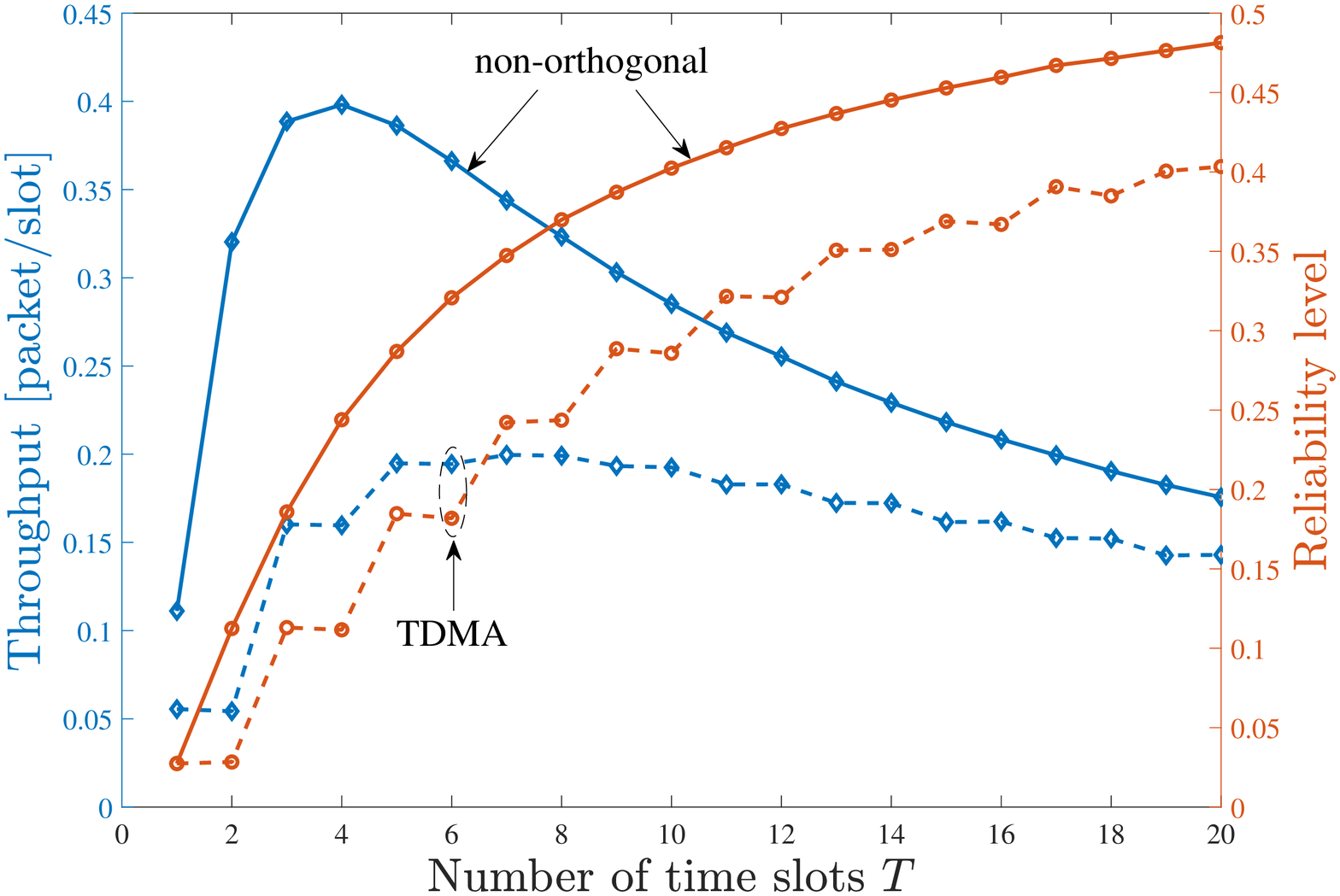}
  \caption{Critical throughput and reliability level}
  \label{fig:throughput_reliability_critical_messages}
\end{subfigure}%
\begin{subfigure}{.5\textwidth}
  \centering
  \includegraphics[height= 5.5 cm, width= 7.5 cm]{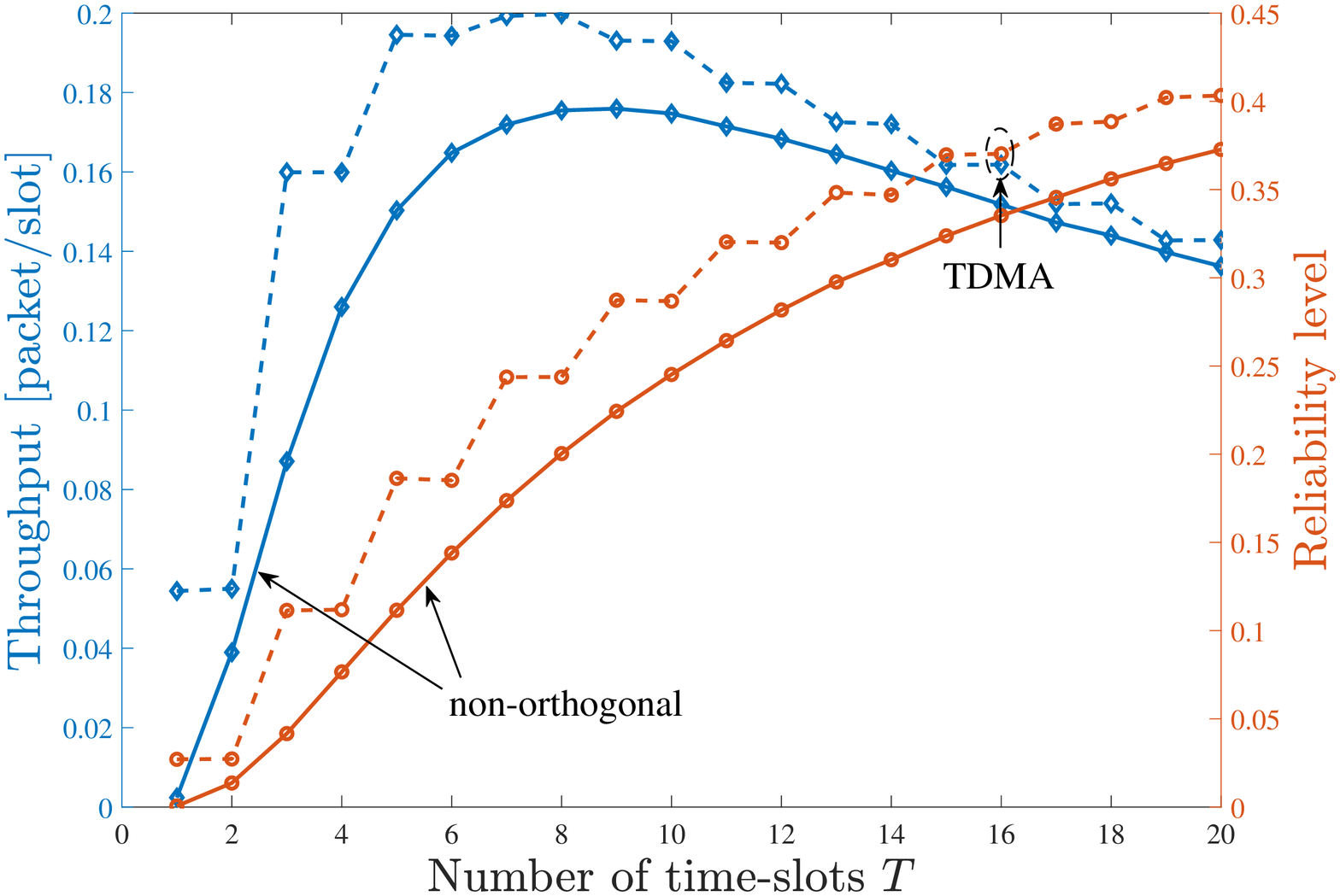}
  \caption{Non-critical throughput and reliability level}
  \label{fig:throughput_reliability_Noncritical_messages}
\end{subfigure}
\caption{Critical and non-critical throughputs and reliability levels as function of the number of time-slots $T$ for non-orthogonal resource allocation (solid lines) and inter-service TDMA (dashed lines) ($G=15\ [\mathrm{packet/frame}]$, $\epsilon_1 = \epsilon_2 = 0.5, L=3$ APs, $\alpha=0.5$ and $\gamma_c = 0.5$).}
\label{fig:throughput_reliability}
\vspace{-10pt}
\end{figure*}
\par Finally, we consider the interplay between the throughputs and reliability levels for both non-orthogonal resource allocation and TDMA as function of the number of time-slots $T$. These are plotted in Fig.~\ref{fig:throughput_reliability} for $G=15\ [\mathrm{packet/frame}]$, $\epsilon_1 = \epsilon_2 = 0.5, L=3$ APs, $\alpha=0.5$ and $\gamma_c = 0.5$. For both services, we observe that the reliability level under both allocation schemes increases as function of $T$. This is because larger value of $T$ decrease chances of packet collisions. However, this not the case for the throughput, since large values of $T$ may cause some time-slots to be left unused, which penalizes the throughput. For the critical service in Fig.~\ref{fig:throughput_reliability_critical_messages}, it is seen that non-orthogonal resource allocation outperforms TDMA in both throughput and reliability level due to the larger number of available resources. Moving to the non-critical service in Fig.~\ref{fig:throughput_reliability_Noncritical_messages}, we observe that TDMA provides better throughput and reliability level than non-orthogonal resource allocation. The main reason for this is that the lower number of resources in TDMA is compensated by the absence of inter-service interference from critical messages. 
\section{Conclusions}
\label{sec:conclusions}
This paper studies grant-free random access for coexisting critical and non-critical services in IoT systems with shared wireless backhaul and uncoordinated access points (APs). A non-orthogonal resource sharing scheme based on random access is considered, whereby critical messages are transmitted with a larger power. From the critical service perspective, it was found that non-orthogonal sharing is preferable to a standard inter-service TDMA protocol in terms of both throughput and reliability level. In contrast, this is not the case for the non-critical service, since inter-service orthogonal resource allocation eliminates interference from the larger-power critical service. Finally, we have identified different regimes in terms of channels erasure probabilities for which increasing the number of APs may be beneficial, thanks to additional space diversity, or harmful, due to the increased inter-AP interference. Among possible extensions of this work, we mention the consideration of a more general collision model in which critical messages can tolerate no more than a given number of interfering non-critical messages \cite{rahif_uncoordinated}.
\bibliographystyle{IEEEtran}
\bibliography{Biblio} 
\end{document}